\documentclass[10pt,amsmath,amssymb,letterpaper]{article}
\usepackage{opex3}
\usepackage{amsmath,amsfonts,amssymb}
\newcommand{\E}{\mathrm{e}}
\newcommand{\I}{\mathrm{i}}
\newcommand{\D}{\mathrm{d}}

\begin{document}

\title{Ptychographic reconstruction of attosecond pulses}

\author{M. Lucchini$^1$, M.H. Br\"ugmann$^2$, A. Ludwig$^1$, L. Gallmann$^{1,2}$, U. Keller$^1$, T. Feurer$^{2,*}$}

\address{$^1$ Department of Physics, ETH Zurich, 8093 Z\"urich, Switzerland \\
$^2$ Institute of Applied Physics, University of Bern, Sidlerstrasse 5, CH-3012 Bern, Switzerland}

\email{$^*$thomas.feurer@iap.unibe.ch}

\begin{abstract}
We demonstrate a new attosecond pulse reconstruction modality which uses an algorithm that is derived from ptychography. In contrast to other methods, energy and delay sampling are not correlated, and as a result, the number of electron spectra to record is considerably smaller. Together with the robust algorithm, this leads to a more precise and fast convergence of the reconstruction.
\end{abstract}


\section{Introduction}

Extracting information from experiments on the dynamics of atoms, molecules or solids, i.e. on the dynamics of their nuclei and electronic charge distribution, is essential for a physical understanding and the verification of theoretical models. The fastest electronic processes, e.g. photo-ionization, Auger decay \cite{Drescher2002}, tunneling ionization \cite{Eckle2008,Landsman2014}, or shake-up \cite{Uibiracker2007}, happen on attosecond time scales. However, before probing of attosecond phenomena became feasible, a number of technical challenges had to be addressed. One of the most pressing challenges was to develop techniques to characterize the electric field of an isolated attosecond pulse or a train of such pulses as they are generated in the process of high-order harmonic generation. Over the past decade attosecond streaking measurements \cite{Hentschel2001} have proven to be powerful techniques in extracting information on attosecond photo-electron emission \cite{Uibiracker2007,Schultze2010,Sabbar2015}. With suitable assumptions they allow for a complete characterization of attosecond pulses \cite{Mairesse2005,Gagnon2009}, making use of iterative algorithms derived from frequency-resolved optical gating (FROG). This technique became known as FROG-CRAB and the shortest pulses characterized to date were 67~as long \cite{Zhao2015}. Recently, an alternative algorithm has been developed which is optimized for attosecond streaking \cite{Gagnon2008,Chini2010} and is especially suitable for pulses approaching the single-cycle limit.

Here, we present a new modality for attosecond pulse characterization which is derived from a phase retrieval scheme widely used in lensless imaging, namely ptychography. It is related to the solution of the phase problem in crystallography as proposed by Hoppe \cite{Hoppe1969} and was first demonstrated experimentally at visible wavelengths \cite{Rodenburg2007}. In ptychography a real space object, in particular its amplitude and phase, is reconstructed iteratively from a series of far-field diffraction measurements. Each of those is recorded after either moving the object or the coherent illumination beam in a plane perpendicular to the propagation direction of the illumination beam. The transverse shift of the illumination beam is smaller than its spatial support, so that subsequent far-field diffraction patterns result from different, but overlapping regions of the object. The spatial resolution is limited by the positioning accuracy, the stability of the entire setup and the angular range of scattered wavevectors that can be recorded with a sufficiently high signal-to-noise ratio. Ptychography has been proven to produce the correct real space image if the illumination beam is known \cite{McCallum1992}, but works even if the illumination beam is unknown in which case its profile is reconstructed together with the object \cite{Thibault2009, Maiden2009}. Applying ptychography to the reconstruction of temporal rather than spatial objects requires operating in one dimension with the conjugated variables time and frequency \cite{Godil1994}. The unknown temporal object is to be reconstructed iteratively from a series of far-field diffraction measurements, i.e. spectra. Each of those is recorded after delaying the coherent illumination pulse with respect to the temporal object with the time delay being smaller than the temporal support of the illumination pulse. The temporal resolution is primarily limited by the range of spectral amplitudes which can be recorded with a sufficiently high signal-to-noise ratio (SNR). Recently, we have shown that ptychography can indeed be applied to reconstruct temporal objects if the illumination pulse is fully characterized \cite{Spangenberg2014}. We have subsequently shown that ptychography is a very powerful technique for ultrafast pulse characterization \cite{Spangenberg2015}. Here, we show that the concept can be extended to attosecond pulse characterization or photo-electron streaking experiments in general.

\section{Methodology}

In attosecond streaking experiments an extreme ultraviolet (XUV) pulse ionizes atoms in the presence of an infrared (IR) laser field \cite{Gallmann2012}, henceforth called streaking field, and the energy spectrum of the photo-electrons released in the process is measured. With no streaking field present and a sufficiently flat ionization cross section the photo-electron spectrum is a copy of the XUV pulse spectrum. When the streaking field overlaps with the emerging electron wave packet the time dependent IR vector potential is imprinted on the photo-electron spectrum. The characterization of the XUV pulse is achieved by measuring the streaked photo-electron spectrum for different time delays between the XUV pulse and the streaking field. Within the single electron approximation the spectrum of the photo-electrons generated by the XUV pulse and modified by the streaking field is

\begin{equation}
\label{eq:electron}
S(p,\tau) = \left| \int \D t \; E(t) \; d(p+A_\mathrm{IR}(t-\tau)) \; P(t-\tau) \; \E^{\I (p^2/2+I_p) t} \right|^2
\end{equation}

with the XUV pulse $E(t)$, the dipole transition matrix element $d(p)$, the gate function $P(t)$, the electron momentum $p$ and the ionization potential $I_p$, all in atomic units. The gate function is a pure phase gate

\begin{equation}
\label{eq:gate}
P(t) = \exp\left[ -\I \int\limits_t^\infty \D t' \left( p_c A_\mathrm{IR}(t') - \frac{A_\mathrm{IR}^2(t')}{2} \right) \right] 
\end{equation}

with the vector potential of the streaking field $A_\mathrm{IR}(t)$ and the unstreaked central momentum $p_c$. The measured electron spectra $S(p,\tau)$ can be readily converted to $S(\omega,\tau)$. When assuming a constant transition matrix element $d$, which is approximately true for not too high streaking fields, Eqn.~(\ref{eq:electron}) is the spectrum of the product field $E(t) P(t-\tau)$. Thus, it should be possible to use the ptychographic iterative engine (PIE) to reconstruct $E(t)$ if $P(t)$ is known \cite{McCallum1992} or the extended ptychographic iterative engine (ePIE) to reconstruct both the XUV pulse as well as the streaking field \cite{Thibault2009, Maiden2009}.

Ptychography operates on two sampling grids which are largely independent from each other. The XUV pulse and the streaking field are sampled on an equidistant temporal grid, with $M$ samples equally spaced by $\delta t$, which is solely determined by the resolution and total spectral range of the spectrometer used --- or by the spectral range which can be detected with a sufficient SNR. The second grid is that of the time delays and consists of $N$ samples equally spaced by $\delta\tau$. Both grids may but do not necessarily have to span the same time window. If they span the same time window the frequency increment of both grids $\delta\nu$ is identical and we find $\delta\nu \delta t = 1/M$ and $\delta\nu \delta\tau = 1/N$, respectively. The only constraint on the two integers $M$ and $N$ is $N \leq M$ but typically $N$ is orders of magnitudes smaller than $M$. Ptychography requires $N$ spectra $I_n(\omega)$ which are recorded at different time delays $\tau_n$ ($n=1 \ldots N$) between the XUV pulse and the streaking field. All spectra combined result in a spectrogram $S(\omega,\tau)$ sampled on an $M \times N$ grid. As a starting point for the reconstruction algorithm we assume white noise for the XUV pulse, i.e. $E_{j=1,n=1}(t)$, and a reasonable gate function $P_{j=1,n=1}(t)$ obtained from the streaking trace itself using the center of mass method \cite{Boge2014}. In every iteration $j$ all measured spectra are processed. For ascending $n$ the algorithm first updates the current estimate of the XUV pulse and hereafter the estimate of the gate function. It calculates the exit field $\xi_{j,n}(t,\tau_n)$ for a particular time delay $\tau_n$ between the gate function $P_{j,n}(t)$ and the XUV pulse $E_{j,n}(t)$

\begin{equation}
\label{eq:xi_E}
\xi_{j,n}(t,\tau_n) = E_{j,n}(t) \; P_{j,n}(t-\tau_n).
\end{equation}

From $\xi_{j,n}(t,\tau_n)$ we calculate the Fourier transform $\xi_{j,n}(\omega,\tau_n)$ and replace its modulus by the square root of the corresponding experimental/simulated spectrum $I_n(\omega)$ while preserving its phase. After an inverse Fourier transformation the new function $\xi'_{j,n}(t,\tau_n)$ differs from the initial estimate and the difference is used to update the current estimate of the XUV pulse

\begin{equation}
\label{eq:update_E}
\nonumber
E_{j,n+1}(t) = E_{j,n}(t) + \beta_E \; U_{j,n}(t-\tau_n) \; [\xi'_{j,n}(t,\tau_n) - \xi_{j,n}(t,\tau_n)]
\end{equation}

with the weight or window function based on the complex conjugate $P_{j,n}^*(t)$

\begin{equation}
\label{eq:Ut}
U_{j,n}(t) = \frac{P_{j,n}^*(t)}{\mathrm{max}\left\{ |P_{j,n}(t)|^2 \right\}}
\end{equation}

and $\beta_E \in \; ]0 \ldots 1]$. Similarly, the algorithm updates the gate function starting from

\begin{equation}
\label{eq:xi_P}
\xi_{j,n}(t,\tau_n) = E_{j,n}(t+\tau_n) \; P_{j,n}(t).
\end{equation}

As before, $\xi_{j,n}(t,\tau_n)$ is Fourier transformed, the modulus is replaced by the square root of the corresponding spectrum $I_n(\omega)$ and the new function $\xi'_{j,n}(t,\tau_n)$, obtained after an inverse Fourier transformation, is used to update the current estimate of the gate function

\begin{equation}
\label{eq:update_P}
\nonumber
P_{j,n+1}(t) = P_{j,n}(t) + \beta_P \; V_{j,n}(t+\tau_n) \; [\xi'_{j,n}(t,\tau_n) - \xi_{j,n}(t,\tau_n)]
\end{equation}

with the weight or window function based on the complex conjugate $E_{j,n}^*(t)$

\begin{equation}
\label{eq:Vt}
V_{j,n}(t) = \frac{E_{j,n}^*(t)}{\mathrm{max}\left\{ |E_{j,n}(t)|^2 \right\}}
\end{equation}

and $\beta_P \in \; ]0 \ldots 1]$. Since the gate function has to be a pure phase gate we can impose an additional constraint, i.e. $P_{j,n+1}(t) = \exp\{ \I \arg[ P_{j,n+1}(t)] \}$. In a typical attosecond streaking experiment the absolute value of the pump-probe time delay is not a priori known but for ePIE the actual value of the time delay axis is meaningful and its zero cannot be arbitrarily defined. This, however, does not constitute a limitation. A shift of time delay zero will result in a shift of the time axis of both reconstructed fields, i.e. $E(t)$ and $P(t)$. Therefore, as long as the absolute value of the time delay at which the streaking trace is centered does not fall outside the total time window, the algorithm converges to the correct but temporally shifted solution.

The ptychographic scheme differs from other attosecond reconstruction modalities, such as the Principal Component Generalized Projections Algorithm (PCGPA) \cite{Kane1999} or the Least-Squares Generalized Projections Algorithm (LSGPA) \cite{Gagnon2008,Chini2010}, specifically: 1) The time delay increment is not related to the desired temporal resolution or the wavelength sampling of the spectrometer, but only to the duration of streaking field $P(t)$. The time delays do not even have to be equidistant. 2) Typically, only a few spectra have to be recorded. 3) The small number of spectra to process and the robust algorithm result in an extremely fast convergence of the retrieval algorithm. The PCGPA algorithm relies on a spectrogram sampled on an $M \times M$ grid that satisfies $\delta\nu \delta t = 1/M$. Typically the spectral axis has a higher sampling rate than the temporal axis, and as a consequence the temporal axis needs to be interpolated. Naturally, interpolation will neglect any temporal structure finer than the original sampling, thus PCGPA cannot accurately reconstruct temporal features which vary more rapidly than the time delay increment. With the LSGPA algorithm this problem is somewhat relaxed to $\delta\nu \delta t = L/M$, with $L \geq 1$ being an integer, as long as two constraints are considered. Firstly, the spectrogram has to be recorded at equidistant time delays (this constraint might be overcome by using a variable $L$ \cite{Gagnon2008}), and secondly, the time delay has to be smaller than the temporal support of the XUV pulse. In contrast to both, the ptychographic scheme has no direct link between temporal sampling and time delay as explained above. The time delay increment $\delta\tau$ is related only to the duration of the slowly varying envelope of the streaking field $P(t)$. The relevant quantity in ptychography is the fundamental sampling ratio. It is defined as the ratio of the full widths at half maximum duration (FWHM) of the streaking field over the time delay increment, i.e. $R = \mathrm{FWHM}[P(t)] / \delta\tau$, and if both are identical the fundamental sampling ratio is equal to one. For a fundamental sampling ratio $R>1$ the streaking field overlaps with parts of the XUV pulse several times and this overlap increases the redundancy in the data recorded. It is well known that this redundancy can be used to not only reconstruct the XUV pulse but also the streaking field \cite{Thibault2009, Maiden2009}. Therefore, we expect attosecond ptychography to work for $R>1$ and indeed we find that $R>5$ gives accurate reconstruction results for both the XUV pulse and the streaking field. Somewhat more challenging is to identify suitable values for $\beta_E$ and $\beta_P$ since no theoretical framework exists which can hint to their optimal values. Through dedicated simulations we can estimate that $\beta_E$ scales approximately inverse with $\sqrt{R}$. For example, a streaking field duration of 5~fs and a time delay increment of 0.2~fs result in $R=25$ and consequently $\beta_E$ should be approximately 0.2. In order to estimate $\beta_P$ we resorted to numerical simulations through which we found that $\beta_P \leq 0.1$ gives excellent results. A small $\beta_P$ value, i.e. a small update rate of the streaking field, does make sense given the fact that the shape of the streaking field is already relatively well defined through the center of mass method \cite{Boge2014}.

\section{Simulations}

\begin{figure}[htb]
\centering
\includegraphics[width=100mm]{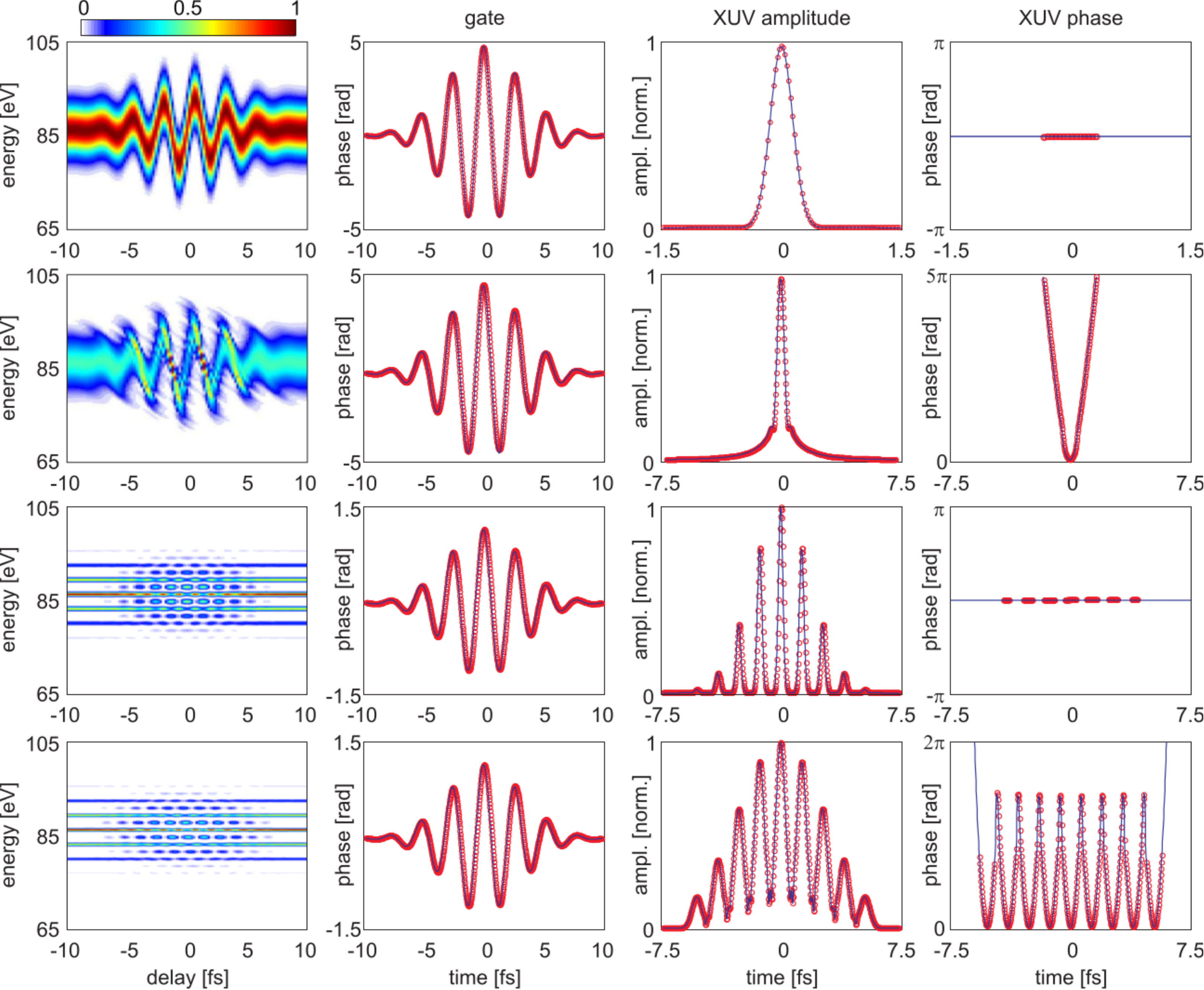}
\caption{\label{fig1} Spectrogram, streaking field, XUV amplitude and phase versus time. Theoretical (blue curves) and ePIE reconstructed (red open circles) results. First row: Isolated bandwidth-limited attosecond pulse of 0.24~fs duration. Second row: Same attosecond pulse with a quartic phase of $2 \cdot 10^{-4}$~fs$^4$. Third row: Pulse train consisting of nine unchirped pulses each 0.24~fs long. Fourth row: Same pulse train with a quadratic phase of $2 \cdot 10^{-2}$~fs$^2$.}
\end{figure}

We first demonstrate that the ePIE algorithm produces accurate reconstruction results for different XUV pulses. We then proceed to identify the optimal reconstruction parameters, i.e. the time delays $\tau_n$, the number of spectra $N$, as well as $\beta_E$ and $\beta_P$. Next we show that only very few spectra are required to obtain accurate reconstruction results and finally we compare the algorithm to  PCGPA and LSGPA and analyze its sensitivity with respect to noise.

\begin{table}[htb]
\begin{center}
\caption{\label{tab1} ePIE parameters for Figs.~\ref{fig1} to \ref{fig4}. Columns 1 to 7 show the figure, row number, number of spectra $N$, time delay increment $\delta\tau$, the ePIE parameters $\beta_E$ and $\beta_P$ and the rms error.}
\begin{tabular}{cc|rrrrr}
\hline
figure & row & $N$ & $\delta\tau$ [fs] & $\beta_E$ & $\beta_P$ & rms \\
\hline
1 & 1 & 101 & 0.20 & 0.35 & 0.10 & $1.8 \cdot 10^{-3}$ \\
  & 2 & 101 & 0.20 & 0.35 & 0.10 & $1.1 \cdot 10^{-3}$ \\
  & 3 & 101 & 0.20 & 0.35 & 0.10 & $5.8 \cdot 10^{-5}$ \\
  & 4 & 101 & 0.20 & 0.35 & 0.10 & $7.6 \cdot 10^{-6}$ \\
\hline
2 & 1 & 401 & 0.05 & 0.35 & 0.10 & $6.5 \cdot 10^{-2}$ \\
  & 2 & 201 & 0.10 & 0.35 & 0.10 & $6.2 \cdot 10^{-3}$ \\
  & 3 &  51 & 0.40 & 0.35 & 0.10 & $3.6 \cdot 10^{-3}$ \\
  & 4 &  25 & 0.80 & 0.35 & 0.10 & $1.0 \cdot 10^{-2}$ \\
\hline
3 & 1 & 401 & 0.05 & 0.15 & 0.10 & $2.4 \cdot 10^{-3}$ \\
  & 2 &  25 & 0.80 & 0.65 & 0.10 & $3.6 \cdot 10^{-3}$ \\
  & 3 &  13 & 1.60 & 0.85 & 0.10 & $4.1 \cdot 10^{-3}$ \\
\hline
4 & 1 &  25 & 0.80 & 0.65 & 0.10 & $3.6 \cdot 10^{-3}$ \\
  & 2 &  25 & 0.40 & 0.55 & 0.10 & $1.7 \cdot 10^{-3}$ \\
  & 3 &  25 & 0.20 & 0.45 & 0.10 & $1.6 \cdot 10^{-3}$ \\
  & 4 &  25 & 0.10 & 0.35 & 0.10 & $6.6 \cdot 10^{-4}$ \\
  & 5 &  25 & 0.05 & 0.25 & 0.10 & $3.6 \cdot 10^{-3}$ \\
\hline
\end{tabular}
\end{center}
\end{table}

In all examples presented hereafter the streaking field is an IR pulse centered at 800~nm. Its spectrum is Gaussian with a FWHM of 180~nm corresponding to a FWHM pulse duration of approximately 5~fs. The pulses are focused to intensities between $10^{11}$~W/cm$^2$ and $10^{13}$~W/cm$^2$ which is sufficiently low to invoke the central momentum approximation. The XUV pulses and the streaking field are sampled on a temporal grid extending $\pm 50$~fs around time zero with a resolution of $\delta t = 0.024$~fs and $M = 2^{12}$. The corresponding energy grid is centered around 86.25~eV with a resolution of 0.042~eV. The transform limited XUV pulse has a FWHM duration of 0.24~fs and we analyze individual pulses with and without phase modulation as well as trains of pulses consisting of two or nine such pulses with different amplitudes and a separation given by one half of the IR driver laser's oscillation period. As a measure for the quality of the reconstruction we use the root mean square (rms) between the original and the reconstructed spectrogram after a finite number of iterations.

\begin{equation}
\mathrm{rms} = \min_\gamma \left\{ \sqrt{\frac{1}{M \; N} \; \sum\limits_{\omega, \tau} \left| S_j(\omega,\tau) - \gamma S(\omega,\tau) \right|^2} \right\}
\end{equation}

with the spectrogram $S_j(\omega,\tau)$ after $j$ iterations calculated from the actual estimate of the XUV pulse and streaking field, and $\gamma$ being a normalization constant that minimizes the difference in intensities between the simulated/measured and the reconstructed spectrograms \cite{Trebino1997}.

\begin{figure}[h!]
\centering
\includegraphics[width=100mm]{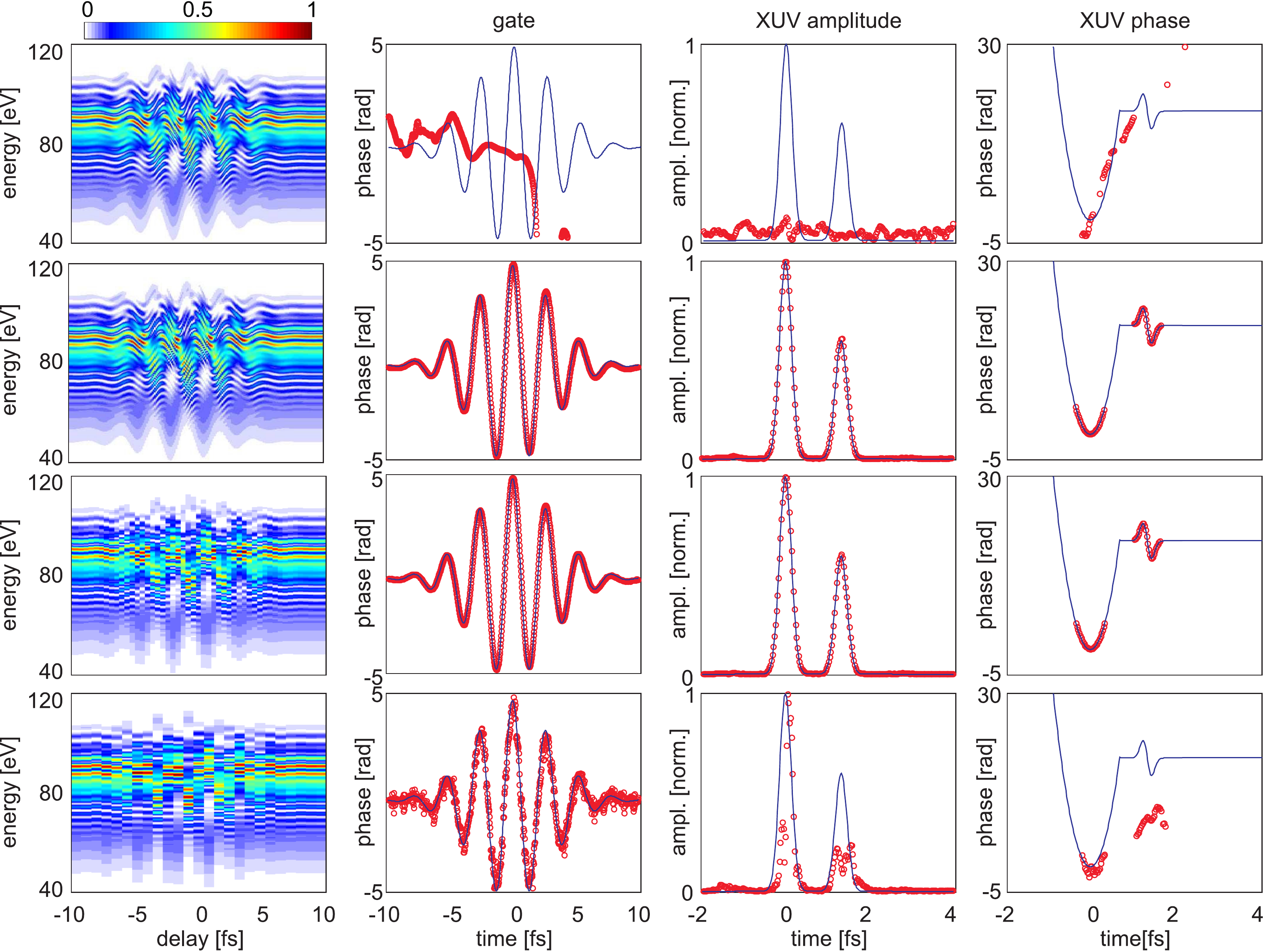}
\caption{\label{fig2} Reconstruction of pulse with satellite with fixed parameters and different numbers of time delays. Spectrogram, streaking field, XUV amplitude and XUV phase. Theoretical (blue curves) and reconstructed (red open circles) results for a double pulse. First row: $N=401$, $\delta\tau = 0.05$~fs. Second row: $N=201$, $\delta\tau = 0.1$~fs. Third row: $N=51$, $\delta\tau = 0.4$~fs. Fourth row: $N=25$, $\delta\tau = 0.8$~fs.}
\end{figure}

\begin{figure}[h!]
\centering
\includegraphics[width=100mm]{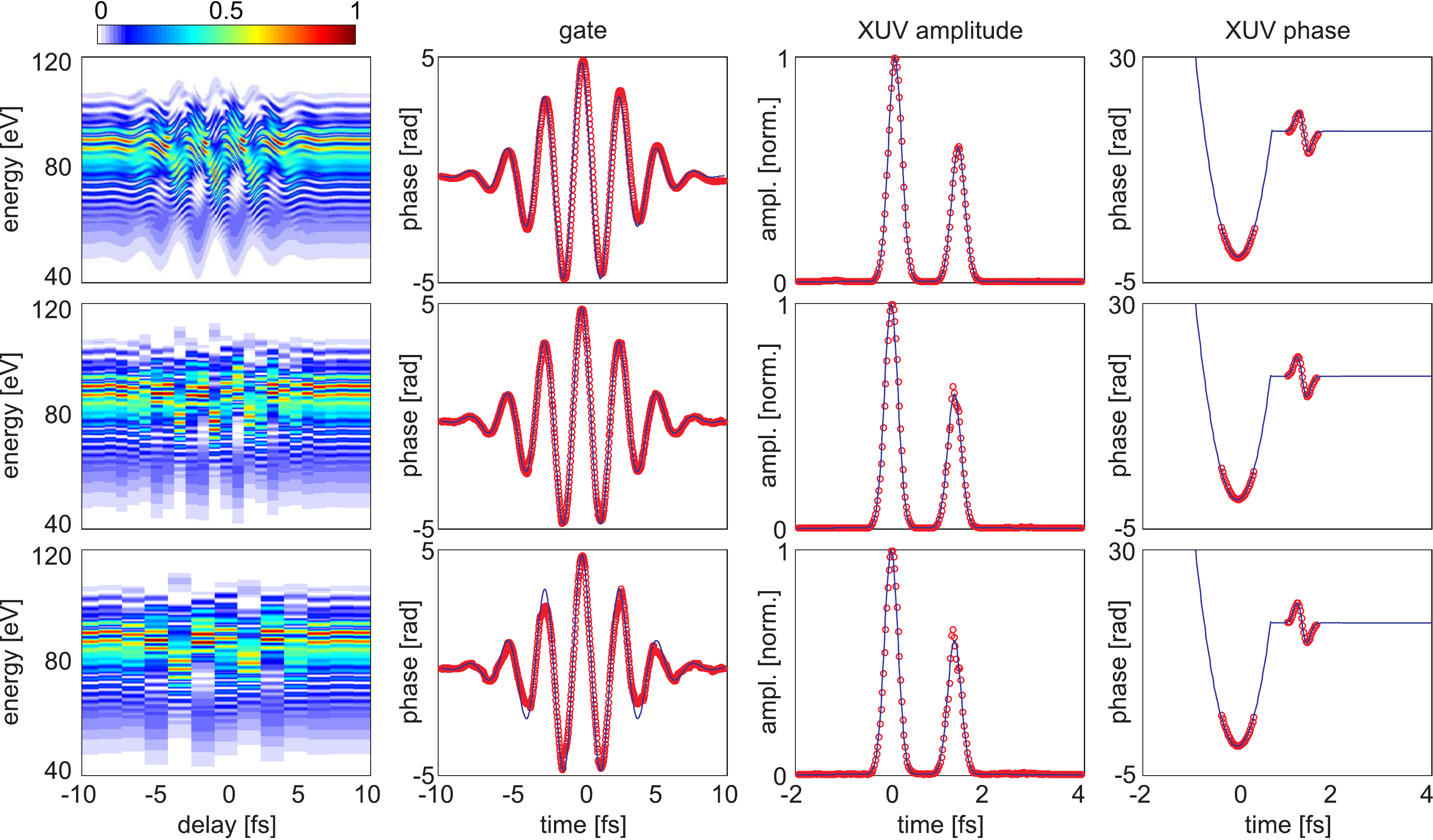}
\caption{\label{fig3} Reconstruction of pulse with satellite with adapted parameters for different numbers of time delays. Spectrogram, streaking field, XUV amplitude and XUV phase. Theoretical (blue curves) and reconstructed (red open circles) results for a double pulse. First row: $N=401$, $\delta\tau = 0.05$~fs, $\beta_E = 0.15$. Second row: $N=25$, $\delta\tau = 0.8$~fs, $\beta_E = 0.65$. Third row: $N=13$, $\delta\tau = 1.6$~fs, $\beta_E = 0.85$.}
\end{figure}

Figure~\ref{fig1} shows the reconstruction results assuming four different XUV pulses. The first row compares simulations (solid blue curve) and reconstruction (red open circles) results for an isolated bandwidth-limited attosecond pulse of 0.24~fs duration. The agreement obtained after 1000 iterations of the ePIE algorithm is excellent. The reconstructed streaking field as well as the reconstructed XUV pulse are practically indistinguishable from the fields used to calculate the spectrogram. The reconstruction parameters used in Fig.~\ref{fig1} and also those of all following figures are summarized in table~\ref{tab1}. The second example (Fig.~\ref{fig1} second row) is the same attosecond pulse with a quartic phase of $2 \cdot 10^{-4}$~fs$^4$. The third and the fourth examples are trains of attosecond pulses without and with a quadratic phase of $2 \cdot 10^{-2}$~fs$^2$. This shows that the ePIE algorithm is able to reconstruct both single attosecond pulses and attosecond pulse trains, independently of their spectral phase. Henceforth we concentrate on a double pulse with identical pulse duration and an intensity ratio of 1:0.6, similar to the one used in reference \cite{Gagnon2008}, in order to show that ePIE can correctly discern between single attosecond pulses and those with a satellite. The phase modulations of the main pulse and the satellite are $40 t^2$ and $-50 (t-T_p/2) \E^{-2 ((t-T_p/2)/0.2)^2}$, with $T_p$ being the oscillation period of the IR driver field and the time $t$ is to be inserted in femtoseconds.

Figure~\ref{fig2} shows the spectrogram and the reconstructed XUV pulse and streaking field for different time delay increments. The number of spectra simulated was adjusted such that the total time window remains constant (20~fs). A too small or too big time increment leads to erroneous reconstruction results. From time domain ptychography it is known that there is a link between the time delay increment $\delta\tau$ and the reconstruction parameter $\beta_E$. Generally, the larger the time delay increment the larger $\beta_E$ has to be chosen in order to obtain convergence and accurate results. Therefore, the unsuccessful reconstructions of the spectrograms shown in Fig.~\ref{fig2} were repeated, but this time with an optimal $\beta_E$ value. The resulting XUV pulses and streaking fields are shown in Fig.~\ref{fig3}. When adjusting $\beta_E$ accordingly, we find excellent agreement for all time delay increments. Note that for $\delta\tau=1.6$~fs (Fig.~\ref{fig3} third row) only 13 spectra are sufficient for an accurate pulse reconstruction.

\begin{figure}[h!]
\centering
\includegraphics[width=100mm]{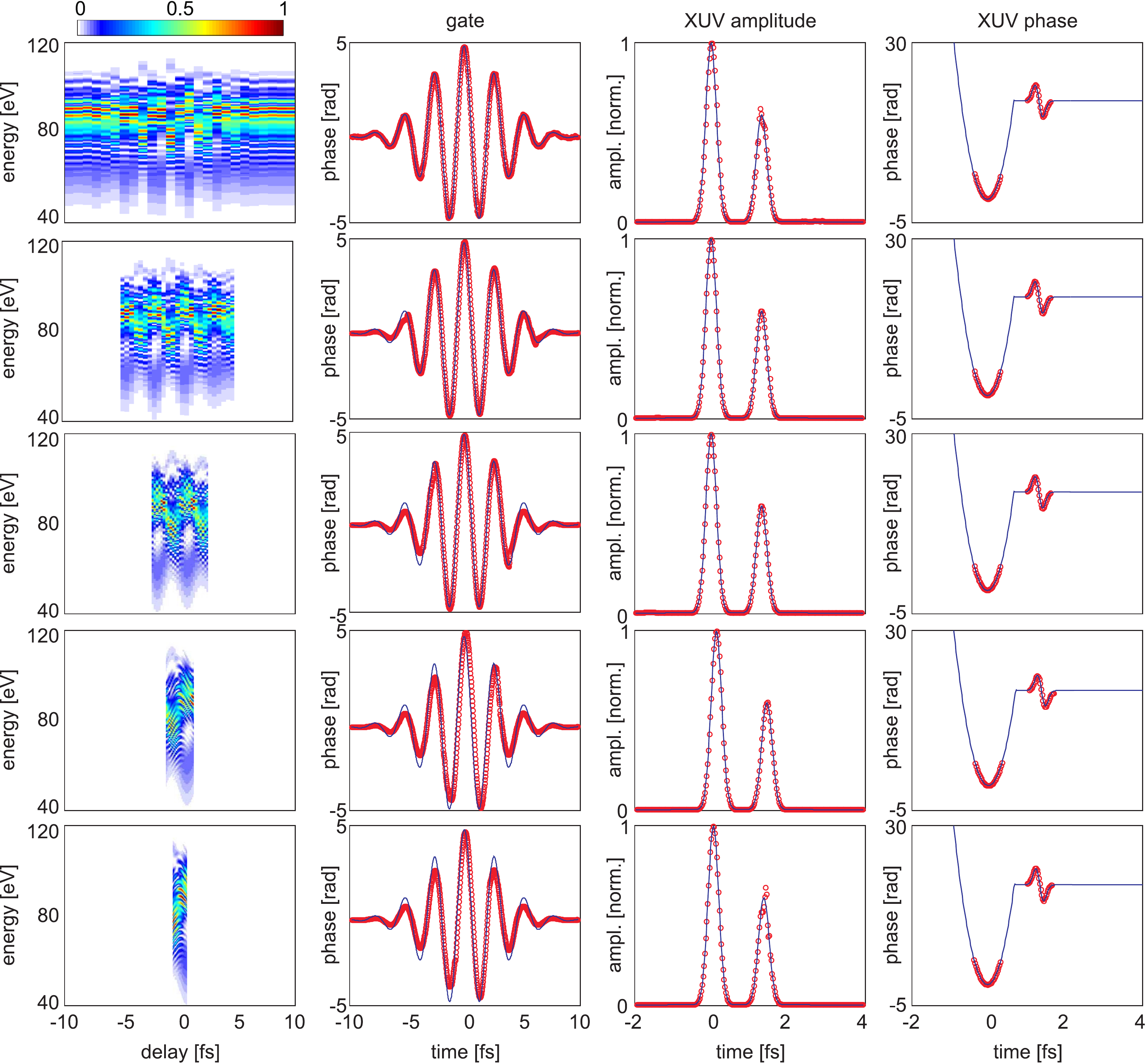}
\caption{\label{fig4} Reconstruction of pulse with satellite with adapted parameters for decreasing coverage of the spectrogram. Spectrogram, streaking field, XUV amplitude and XUV phase. Theoretical (blue curves) and reconstructed (red open circles) results for a double pulse. First row: $\delta\tau = 0.8$~fs, $\beta_E = 0.65$. Second row: $\delta\tau = 0.4$~fs, $\beta_E = 0.55$. Third row: $\delta\tau = 0.2$~fs, $\beta_E = 0.45$. Fourth row: $\delta\tau = 0.1$~fs, $\beta_E = 0.35$. Fifth row: $\delta\tau = 0.05$~fs, $\beta_E = 0.25$.}
\end{figure}

In the following we explore the size of the streak window required for a sufficiently accurate reconstruction result. For this purpose we fix the number of spectra and decrease the time delay step from 0.8~fs to 0.05~fs. Figure~\ref{fig4} shows the reconstruction results. For 0.05~fs we cover a total streak window of only 1.25~fs which corresponds to roughly one half of the streaking field period and is comparable to the separation between the two attosecond pulses. Despite the small streak window, which just about extends over the double pulse waveform, the reconstruction of the XUV pulse works surprisingly well. For isolated attosecond pulses an even shorter total streak window may be sufficient. The PCGPA algorithm, requiring periodic boundary conditions, fails to converge if the streaking field is only partially covered by the range of time delays. In LSGPA the problem is somewhat relaxed. In stark contrast, the ePIE algorithm only requires a range of time delays that covers the XUV pulse to be reconstructed but not the entire streaking field. Of course the reconstruction of the streaking field outside the streak window is arbitrary and without relevance.

\begin{figure}[h!]
\centering
\includegraphics[width=100mm]{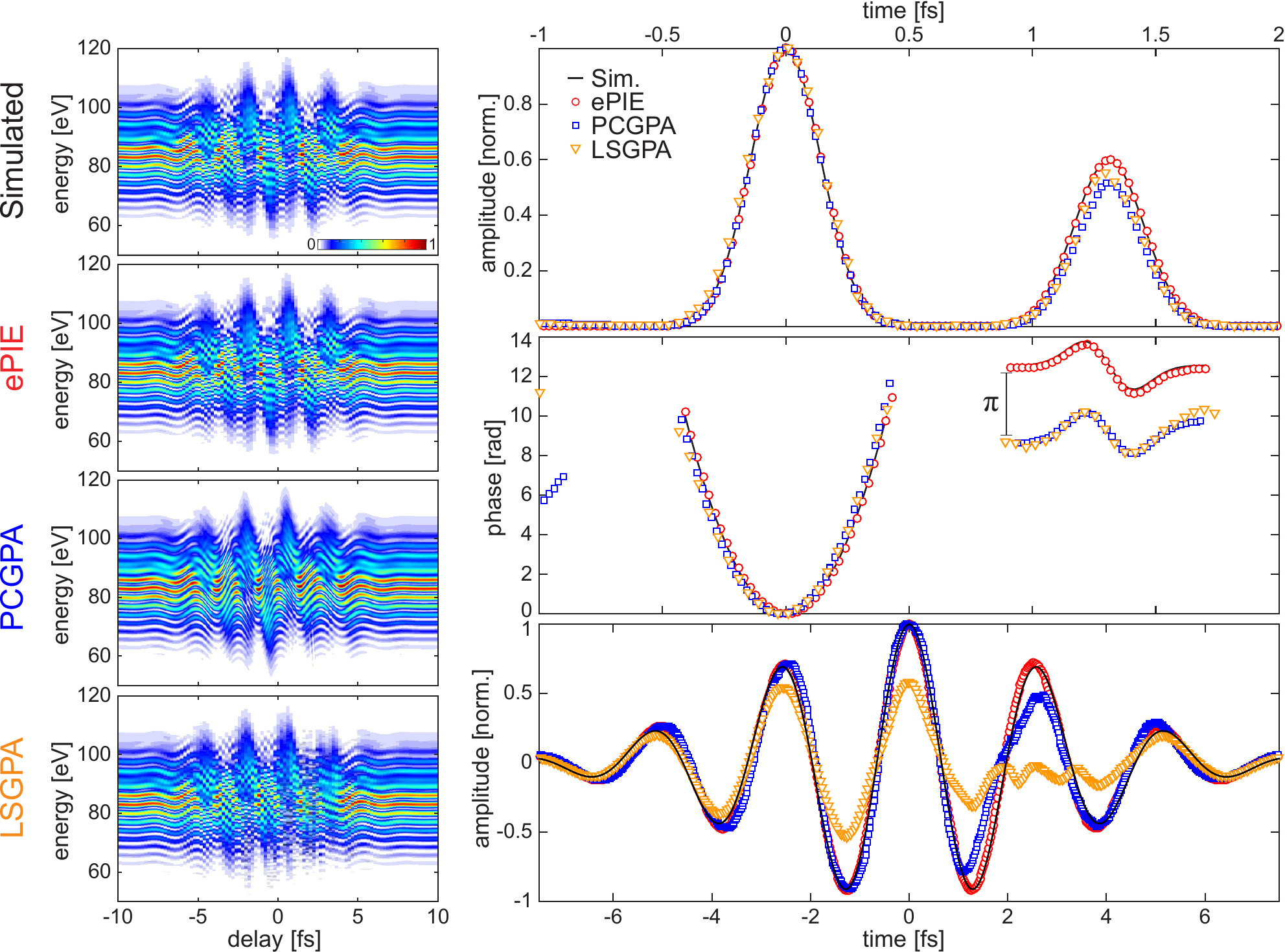}
\caption{\label{fig5} Comparison of ePIE with PCGAP and LSGPA for the double attosecond pulse. Left column: Simulated and reconstructed spectrograms; Right column: Retrieved time dependent amplitude and phase of the XUV pulse and the streaking field. First top-left panel and the black solid curves: Simulated data. Reconstructed pulses in red circles, blue squares and orange triangles for ePIE, PCGPA and LSGPA. Each reconstruction is after 20000 iterations, resulting in final errors of $1.4 \cdot 10^{-4}$, $2.5 \cdot 10^{-2}$ and $3.4 \cdot 10^{-2}$ for ePIE, PCGPA and LSGPA.}
\end{figure}

Next we investigate ePIE's performance in comparison to other algorithms commonly used in attosecond streaking, i.e. PCGPA or LSGPA. Figure~\ref{fig5} compares the results of PCGPA and LSGPA to those of ePIE for the case of the double attosecond pulse with a phase modulation of $60 t^2$ for the main pulse and $-20 (t-T_p/2) \E^{-2 ((t-T_p/2)/0.2)^2}$ for the satellite. The IR streaking field intensity is $5.5 \cdot 10^{12}$~W/cm$^2$. The simulated spectrogram (Fig.~\ref{fig5} top-left panel) is composed of $N=105$ spectra which are simulated with a time delay increment of $\delta\tau = 0.2$~fs. The black curves in the right column of the Fig.~\ref{fig5} show the amplitude (top) and phase (center) of the simulated XUV pulse. The lower panel displays the IR streaking field. The reconstruction results presented are after 20000 iterations for all three algorithms. The second spectrogram from the top shows the ePIE reconstruction. The extracted XUV pulse and streaking field are marked with red circles in the respective panels. The third spectrogram from the top shows the PCGPA results and the bottom one the reconstruction of the LSGPA (corresponding to blue open squares and orange open triangles in the right column). For the case of ePIE the entire generated spectrogram is used in the reconstruction algorithm. In order to keep the computation time reasonable, the spectrogram as been reduced in size to a square matrix composed of $2^{10} \times 2^{10}$ points for PCGPA. In the case of LSGPA only the energy axis has been reduced to $2^{10}$ points which results in a delay-energy step ratio of $L = 5$. By comparing the different results it is possible to conclude that, while all the algorithms correctly reconstruct the main pulse, only the ePIE algorithm properly resolves the satellite and the IR field. Moreover, there are a number of marked advantages using ePIE. First, convergence is usually much faster and takes less time as the computational effort is smaller. Second, ePIE produces accurate results even for very few spectra recorded as already discussed above while the two other algorithms fail to converge.

\begin{figure}[htb]
\centering
\includegraphics[width=100mm]{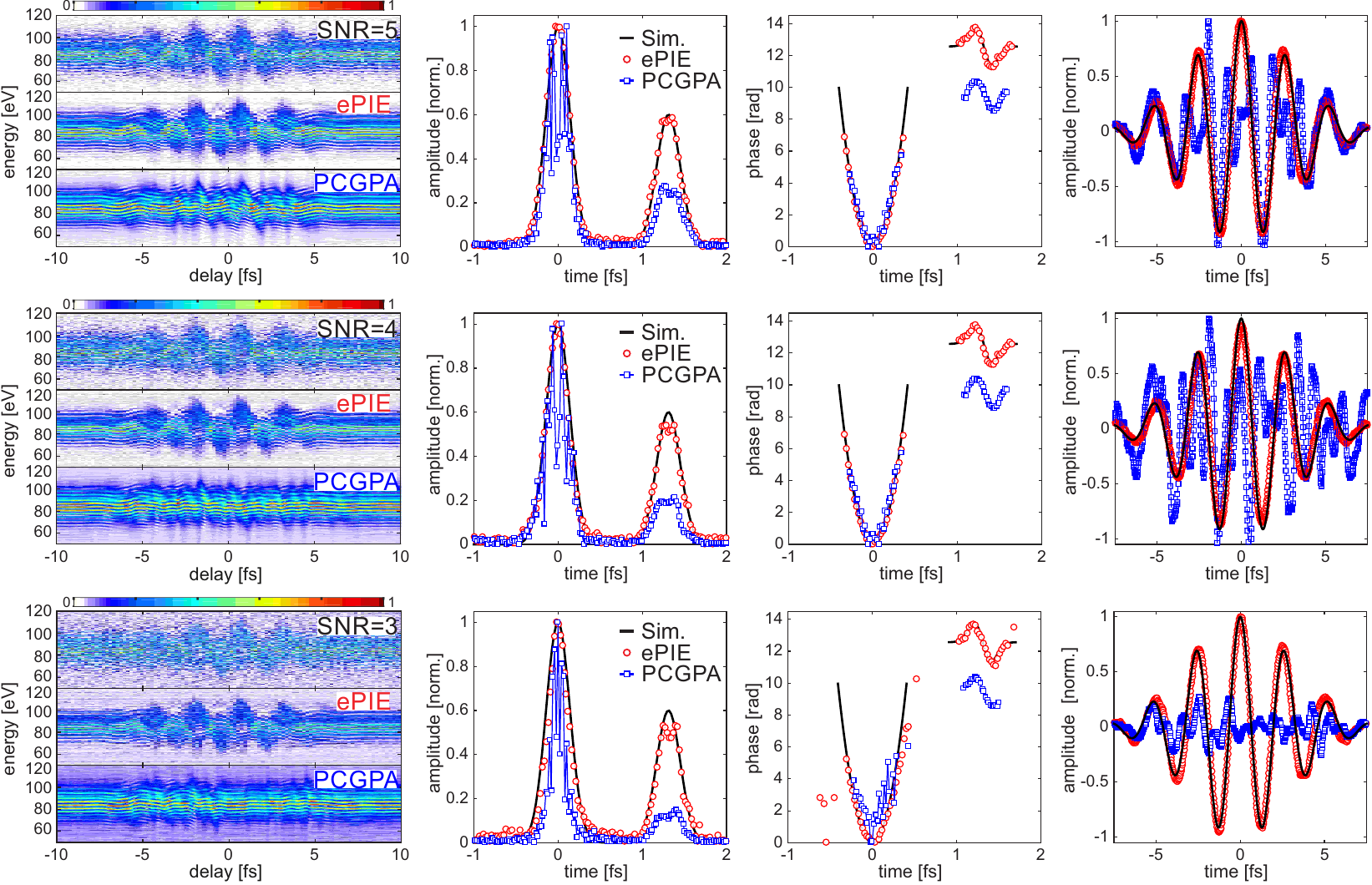}
\caption{\label{fig6} Comparison between ePIE and PCGPA reconstructions in the presence of noise. The first two rows show the data for $\mathrm{SNR} = 5$. The third and fourth rows the case of $\mathrm{SNR} = 4$ and the last two the case of $\mathrm{SNR} = 3$. The simulated XUV and IR pulses are in black curves. Blue squares and red circles display the PCGPA and ePIE reconstruction results.}
\end{figure}

Lastly, we analyze the reconstruction results in the presence of noise. For that purpose we add white noise to the simulated trace in the top-left panel of Fig.~\ref{fig5} prior to the reconstruction. Specifically, we add to each pixel of the spectrogram, which we assume is normalized to a maximum of one, $\mathrm{rand}(-1,1)/\mathrm{SNR}$, with $\mathrm{rand}(-1,1)$ being a random number with Gaussian distribution between $-1$ and $1$. Figure~\ref{fig6} shows three examples with a SNR of 5, 4 and 3. For a $\mathrm{SNR} = 5$ the ePIE algorithm successfully reconstructs both the double pulse and the IR streaking field while the PCGPA results are already very noisy. For a SNR of 4, also the reconstruction of the ePIE starts to become noisy and to underestimate the amplitude of the satellite. Nevertheless, the results are still reasonable since the main characteristics of both XUV and IR are retrieved. With a $\mathrm{SNR} = 3$ the ePIE algorithm still converges and gives comparable results. Only for a SNR as low as 2 the convergence is severely compromised. Note that for the example presented in Fig.~\ref{fig6} the streaking trace is reconstructed without the need for any further interpolation, while in the case of PCGPA the spectrogram has to be interpolated to a square matrix of $2^{10} \times 2^{10}$ points.

\section{Experiment}

\begin{figure}[htb]
\centering
\includegraphics[width=100mm]{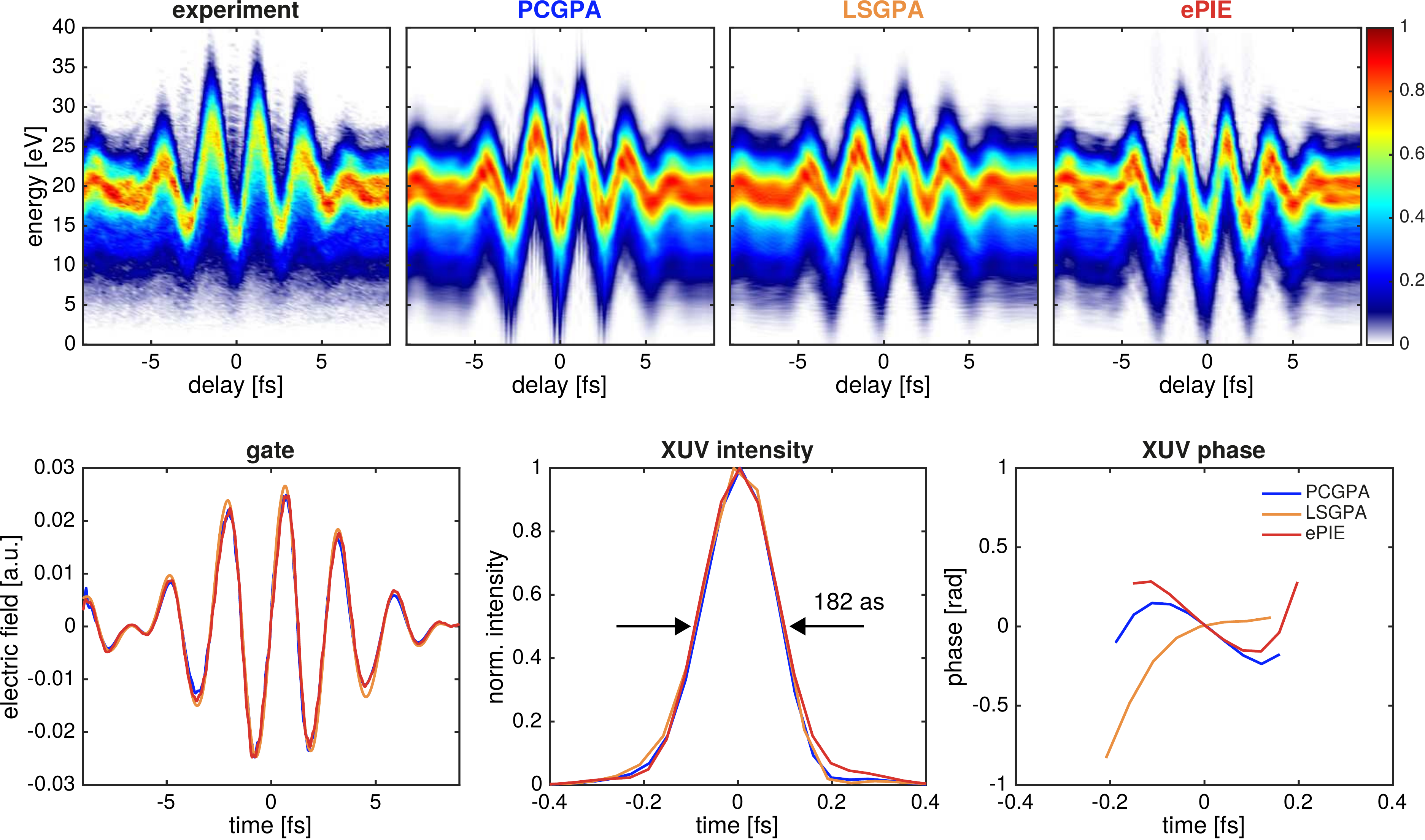}
\caption{\label{fig7} Comparison of reconstruction between PCGPA, LSGPA and ePIE applied to an experimental streaking trace. Top row: Reconstructed spectrograms after 20000 iterations for PCGPA, LSGPA and ePIE. Lower panels: retrieved streaking fields and XUV pulse intensities and phases.}
\end{figure}

We conclude with the ePIE reconstruction of an experimental spectrogram and compare the results with the PCGPA and LSGPA algorithms. The experimental data was acquired at the attoline at ETH Zurich \cite{Locher2014} by ionization of neon with single attosecond XUV pulses which were generated in an argon gas target. Consecutive streaking of the resulting electrons was induced by a co-propagating near-infrared field with a peak intensity of $6.6 \cdot 10^{12}$~W/cm$^2$. The photo-electron energy was measured with a time-of-flight spectrometer for different time delays between the XUV pulse and the streaking field at increments of 0.2~fs. The resulting spectrogram was used to reconstruct the amplitude and the phase of the XUV pulse as well as the streaking field. The PCGPA algorithm requires an interpolation of the experimental spectrogram along the time delay axis prior to reconstruction. We therefore interpolate the spectrogram on a square grid of size $512 \times 512$ fulfilling the Fourier condition $\delta\nu \delta t = 1/M$. It is known that this interpolation can lead to artifacts in the reconstruction \cite{Gagnon2009}. The LSGPA ($L = 4$) as well as ePIE do not require an interpolation and operate on a grid of $512 \times 100$. The results after 20000 iterations are shown in Fig.~\ref{fig7}. The rms errors of PCGPA, LSGPA, and ePIE are $3.7 \cdot 10^{-2}$, $5.6 \cdot 10^{-2}$, and $4.6 \cdot 10^{-2}$, respectively. The reconstructed streaking fields, the XUV intensities with a duration of 182~as and the XUV phases are almost identical. Due to the center of mass approximation applied in all algorithms, the reconstructions cannot account for the asymmetric streaking of the experimental trace. The ePIE reconstruction results are virtually identical to those of PCGPA but require considerably less computational effort due to the lack of constraints on the time delay axis. Thus, we demonstrated that ePIE is suitable for an accurate reconstruction of experimental spectrograms and the characterization of single attosecond pulses.

\section{Conclusion}

In conclusion, we have demonstrated a new modality for attosecond pulse reconstruction based on a retrieval algorithm derived from ptychography which requires to record only a small number of spectra and converges extremely fast and reliably. It reconstructs single attosecond pulses as well as trains of attosecond pulses with an unprecedented degree of accuracy. With a judicious choice of the reconstruction constants it works for a large range of time delay increments and a surprisingly small amount of data. In contrast to all other algorithms, the range of time delays needs to extend only over the temporal support of the XUV pulse but not the streaking field. Of course this is true only if the exact shape of the streaking field is irrelevant beyond the extent of the XUV pulse. The ePIE generally performs superior to algorithms based on general projections, requires considerable less computational effort and is much less susceptible to noise.

\section*{Acknowledgments}

We gratefully acknowledge financial support from the NCCR MUST research instrument of the Swiss National Science Foundation.

\end{document}